\documentclass{ifacconf}

\usepackage{graphicx}      % include this line if your document contains figures
\usepackage{natbib,xcolor}        % required for bibliography

\let\theoremstyle\relax
\usepackage{amsthm}
\usepackage{amsmath,amssymb,amsfonts}

\usepackage{booktabs}
\usepackage{diagbox}
\newtheorem{theorem}{Theorem}
\newtheorem{lemma}{Lemma}

\newtheorem{corollary}{Corollary}

\newtheorem{assumption}{Assumption}

\theoremstyle{remark}

\begin{document}
\begin{frontmatter}

\title{Distributed Control of Network Systems in the Space of Stabilizing Graph Neural Network Policies} 

\author[First]{John Cao} 
\author[First]{Luca Furieri} 

\address[First]{Department of Engineering Science, University of Oxford (e-mail: john.cao@oriel.ox.ac.uk, luca.furieri@eng.ox.ac.uk).}

\begin{abstract}
We study distributed control of networked systems through reinforcement learning, where neural policies must be simultaneously scalable, expressive and stabilizing. We introduce a policy parameterization that embeds Graph Neural Networks (GNNs) into a Youla-like magnitude-direction parameterization, yielding distributed stochastic controllers that guarantee network-level closed-loop stability by design. The magnitude is implemented as a stable operator consisting of a GNN acting on disturbance feedback, while the direction is a GNN acting on local observations. We prove robustness of the policy to perturbations in both the graph topology and model parameters. Numerical experiments validate the effectiveness of the proposed approach.
\end{abstract}

\begin{keyword}
Learning methods for optimal control; Reinforcement learning; Distributed robust controller synthesis; Stability of nonlinear systems; Graph neural networks
\end{keyword}

\end{frontmatter}
%===============================================================================

\section{Introduction}
\vspace{-0.3cm}
The control of large-scale dynamical systems is a fundamental topic in control theory, with applications ranging from UAV coordination \citep{qie2019joint}, search and rescue \citep{kumar2004robot} to warehouse robotics \citep{kattepur2018distributed}. Learning-based controllers often outperform traditional controllers on complex systems. In distributed settings, they are typically implemented under the \emph{centralized training with decentralized execution} (CTDE) paradigm, often using actor-critic methods \citep{konda1999actor} with a centralized critic and decentralized actor. Despite strong empirical results in multi-agent reinforcement learning (MARL), CTDE methods face major challenges: lack of formal stability guarantees, high training variance, and poor generalization when the number of agents changes between training and deployment. Despite its importance in safety-critical applications, methods that ensure stability while retaining expressivity remain underexplored. 

Recent approaches often enforce stability using simple nominal controllers. \cite{zhang2025gcbf+} consider a multi-agent navigation problem where LQR- and PID-based controllers heuristically stabilize the system with a learned safety filter added on top. However, convergence is not guaranteed. Combined with safety filtering, basic nominal policies can also induce jittery, bang-bang inputs \citep{borquez2024safety}. Methods using neural networks generally lack stability guarantees and can destabilize the system unless carefully tuned \citep{fazel2018global}. These limitations highlight the need for policies that are both expressive and scalable while retaining stability guarantees.
% Designing such policies is a challenging task, as distributed optimal control is hard even for linear quadratic (LQ) cases, where optimization becomes impractical due to the exponential growth of the search space. For nonlinear networked systems, deriving performant distributed stabilizing controllers is substantially more difficult, which motivates the need for expressive neural policies. However, direct neural parameterizations generally lack stability guarantees and can destabilize the system unless carefully tuned \citep{fazel2018global}.

Several works aim to address these problems by incorporating stability directly into the synthesis of neural policies. \cite{lale2022kcrl} learn stabilizing policies using a Lyapunov constraint derived from Krasovskii’s method. An adjacent line of work \citep{roberts2011feedback, furieri2022neural, furieri2024learning} generalizes Youla and system-level parameterizations to nonlinear systems, characterizing all stabilizing policies through the lens of disturbance feedback. These methods, however, are either limited by restrictive assumptions on system and controller structures, or rely exclusively on disturbance feedback without exploiting state and graph-structured information, limiting their expressivity and applicability. Recently, \cite{furieri2025mad} extended these policies to include explicit state feedback. However, the method focuses on fully centralized setups, limiting its scalability and ability to transfer to systems of varying sizes.

In parallel, distributed neural control has been studied for structured systems. \cite{furieri2022distributed} learn large-scale stabilizing distributed controllers for port-Hamiltonian networks, while \cite{saccani2024optimal} enforce stability during training of neural controllers in networked settings. However, tailored assumptions on system structure often limit their general applicability, and they do not address scalability in the sense of transferring between varying system sizes. \cite{gama2022distributed} address these challenges by using Graph Neural Networks (GNNs) as deterministic controllers for LQR problems, enabling training on small networks and generalize to larger ones. However, stability and robustness guarantees are established only for linear systems and small GNN parameters. Despite these advances, a unifying framework that simultaneously offers (1) distributed computation with guaranteed closed-loop stability, (2) robustness to topology changes with transfer from small to large systems, and (3) state-of-the-art performance remains a desideratum.

We make the following contributions:

\begin{itemize}
    \item \textbf{Distributed and scalable GNN-based policies:} We present a Youla-like distributed policy parametrization that guarantees closed-loop stability while being realized by GNNs, enabling training on small networked systems and transferring to larger ones with different topologies during deployment.
    
    \item \textbf{Integration with PPO:} We provide a recipe for making our parameterization compatible with Proximal Policy Optimization \citep{schulman2017proximal}.
    
    \item \textbf{Empirical validation of benefits:} We show that our method trained on a few agents generalize to larger systems with unseen network topologies, and highlight the benefits of our method through carefully selected ablations.
\end{itemize}
\vspace{-0.1cm}
\subsection*{Notation}
We denote by $\ell^n$ the set of all sequences $\mathbf{x} = (x_0, x_1, \ldots)$ where $\mathbf{x}_t \in \mathbb{R}^n$. For $p \in \mathbb{N}$, we write $\mathbf{x} \in \ell_p^n$ if $\|\mathbf{x}\|_p = \left( \sum_{t=0}^\infty |\mathbf{x}_t|^p \right)^{1/p} < \infty$, where $|\cdot|$ denotes any vector norm. When clear, we omit the superscript from $\ell^n$ (resp. $\ell_p^n$) and write $\ell$ (resp. $\ell_p$). For $\mathbf{x} \in \ell_p$, we denote its truncation by $\mathbf{x}_{j:i} = (x_i, x_{i+1},\dots,x_j)$. An operator $\mathbf{A} : \ell^n_p \to \ell^m_p$ is \emph{causal} if $\mathbf{A}(\mathbf{x}) = \big( A_0(x_0), A_1(x_{1:0}), \dots, A_t(x_{t:0}), \dots \big)$, and \emph{strictly causal} if $A_t(x_{t:0}) = A_t(0,x_{t-1:0})$. We also write $A_{j:i}(x_{j:0}) = \big( A_i(x_{i:0}), \dots, A_j(x_{j:0}) \big)$. An operator is \emph{$\ell_p$-stable} if it is causal and $\mathbf{A}(\mathbf{a}) \in \ell_p^m$ for all $\mathbf{a} \in \ell_p^n$, and we denote this as $\mathbf{A} \in \mathcal{L}_p$. Finally, an $\mathcal{L}_p$ operator $\mathbf{A} : \mathbf{w} \mapsto \mathbf{u}$ has finite $\mathcal{L}_p$-gain $\gamma(\mathbf{A}) > 0$ if $\|\mathbf{u}\|_p \le \gamma(\mathbf{A})\, \|\mathbf{w}\|_p$ for all $\mathbf{w} \in \ell_p^n$.
Given an undirected graph~$\mathcal{G}=(\mathcal{V}, \mathcal{E})$ described by the set of nodes $\mathcal{V}$ and the set of edges $\mathcal{E}\subset \mathcal{V} \times \mathcal{V}$, we denote the set of neighbors of node~$i$, including~$i$ itself by $\mathcal{N}_i = \{i\} \cup \{j\ |\ \{i,j\}\in\mathcal{E}\} \subseteq \mathcal{V}$. We denote with col$_{j\in\mathcal{V}}(v^{[j]})$ a vector which consists of the stacked
subvectors $v^{[j]}$ from $j=1$ to $j=|\mathcal{V}|$ and with $v^{[\mathcal{N}_i]}$ a vector composed by the stacked subvectors $v^{[i]}$, i.e. $v^{[\mathcal{N}_i]}=col_{i\in\mathcal{N}_i}(v^{[i]})$.

\section{Problem Formulation}
\subsection{Distributed Nonlinear Systems}
We consider a network of $N$ interconnected nonlinear subsystems. The coupling network among the subsystems is defined as an undirected dynamical coupling graph $\mathcal{G}_d=(\mathcal{V}_d, \mathcal{E}_d)$, denoted by the $d$ subscript, with $\mathcal{V}_d=\{1, \dots, N\}$ representing the subsystems in the network, and the set of edges~$\mathcal{E}_d$ containing the pairs of subsystems~$\{i,j\}$, which communicate with each other. Each subsystem is denoted
\begin{equation}
    x_t^{[i]}=f^{[i]}(x^{[\mathcal{N}_{i,d}]}_{t-1},u^{[i]}_{t-1})+w^{[i]}_t,  \ \ \   \forall t\geq1, \label{eq:nonlinsyslocal}
\end{equation}
where state and input of each subsystem $i\in\mathcal{V}_d$ are denoted by $x^{[i]}\in\mathbb{R}^{n_i}$ and $u^{[i]} \in \mathbb{R}^{m_i}$ respectively and $w^{[i]} \in \mathbb{R}^{n_i}$ is an unknown process noise with $w^{[i]}_0=x^{[i]}_0$.\\
In operator form, we can express the subsystem~\eqref{eq:nonlinsyslocal} as:
\begin{equation} \label{eq:sub-operator}
    \mathbf{x}^{[i]} =\mathbf{F}^{[i]}(\mathbf{x}^{[\mathcal{N}_{i,d}]},\mathbf{u}^{[i]})+\mathbf{w}^{[i]},
\end{equation}
where $\mathbf{F}^{[i]}:\ell^{n_{\mathcal{N}_{i,d}}}\times \ell^{m_i} \rightarrow \ell^{n_i}$ is a strictly causal operator defining the dynamics such that $\mathbf{F^{[i]}}(\mathbf{x^{[\mathcal{N}_{i,d}]},u^{[i]}}) $$=\left(0,f^{[i]}\left(x^{[\mathcal{N}_{i,d}]}_0,u_0^{[i]}\right),\dots,f^{[i]}\left(x^{[\mathcal{N}_{i,d}]}_t,u^{[i]}_t\right),\dots\right)$.\\
By combining the local system dynamics in~\eqref{eq:nonlinsyslocal}, the dynamics of the global system result in
\begin{equation} 
    x_t=f(x_{t-1},u_{t-1})+w_t,  \ \ \   \forall t\geq1, \label{eq:nonlinsys}
\end{equation}
where $x_t=col_{i\in\mathcal{V}}(x^{[i]}_t)\in\mathbb{R}^{n}$, $u_t=col_{i\in\mathcal{V}}(u^{[i]}_t)\in\mathbb{R}^{m}$,$w_t=col_{i\in\mathcal{V}}(w^{[i]}_t)\in\mathbb{R}^{n}$.
Similarly to the subsystems, we can rewrite system~\eqref{eq:nonlinsys} in operator form as:
\begin{equation} \label{eq:operator}
    \mathbf{x} =\mathbf{F}(\mathbf{x,u})+\mathbf{w}.
\end{equation}
Every pair of disturbance and input sequences maps to a corresponding state trajectory, inducing a causal operator $\boldsymbol{\mathcal{F}}: (\mathbf{u}, \mathbf{w}) \rightarrow \mathbf{x}$ mapping inputs and disturbances to trajectories. We make the following Assumption on the system to be controlled:
\begin{assumption}
\label{ass:open_loop stability}
    We assume that $\boldsymbol{\mathcal{F}}$ is such that the map $(\mathbf{u},\mathbf{w})\mapsto \mathbf{x}$ lies in $\mathcal{L}_p$, and that the process noise $w_t\sim \mathcal{D}$ is distributed according to an unknown distribution $\mathcal{D}$, and that $\mathbf{w}$ belongs to $\ell^n_p$.
\end{assumption}

The Assumption above means that the interconnection is stable or locally controlled to achieve $\mathcal{L}_p$ stability. 
This is typically true in practical applications where systems are either pre-stabilized or can be stabilized using simple local controllers. While these controllers ensure stability, they often result in suboptimal closed-loop performance. Our goal is to enhance performance while maintaining stability in the space of distributed controllers.
\subsection{Distributed Optimal Control Over Stochastic Policies}
With this formalism, and under Assumption~\ref{ass:open_loop stability}, we define the following \textit{distributed stability constrained optimal control} over dynamic stochastic policies as follows:
\begin{align}
    &\max_\theta \text{ } \mathbb{E}_{\pi_\theta, w_t \sim \mathcal{D}}\left [\sum_{t=0}^\infty \gamma^tR(x_t, u_t) \right ] \label{rl_formluation} \\
    &\text{s.t} \hspace{0.5cm} \mathbf{x} = \boldsymbol{\mathcal{F}}(\mathbf{u}, \mathbf{w}), \notag \\
    &\hspace{0.92cm} u^{[i]}_t \sim \pi^{[i]}_\theta\left(\cdot \text{ } |~x^{[\mathcal{N}^{i,K}_{\mathcal{G}_c}]}_{t:0}, u^{[\mathcal{N}^{i,K}_{\mathcal{G}_c}]}_{t-1:0}, s_{t:0}^{[i]}\right),\quad \forall i \in [1,N]\,,\label{eq:distributed_requirement}\\%= \prod_{i=1}^N \pi^{(i)}_\theta(x_t^{(i)}, w_t^{(i)} | \mathcal{N}_t(i)), \notag \\
    &\hspace{0.92cm} \mathbf{u} \in \ell_p,~\forall \mathbf{w} \in \ell_p\label{eq:stability_constraint}\,.
    %&\hspace{0.92cm} \pi_\theta \in \mathcal{L}_p
\end{align}

The constraint \eqref{eq:stability_constraint} together with Assumption \ref{ass:open_loop stability} enforces closed-loop stability. In particular, restricting $\mathbf{u} \in \ell_p$ whenever $\mathbf{w} \in \ell_p$ implies that the induced trajectory $\mathbf{x}$ lies in $\ell_p$ through the stability of the operator $\boldsymbol{\mathcal{F}}$. The distributed nature of the controller is encoded in \eqref{eq:distributed_requirement}, where $\mathcal{G}_c = (\mathcal{V}_c, \mathcal{E}_c)$ is a \textit{communication graph} and $\mathcal{N}_{\mathcal{G}_c}^{i,K} = \{j \in \mathcal{V}_c | \text{dist}(i,j) \leq K\}$ denotes the $K$-hop neighbors of node $i$ in the graph, and $s_{t:0}^{[i]}$ is a internal controller state for node $i$. Unlike $\mathcal{G}_d$, the graph $\mathcal{G}_c$ might be independent of the system dynamics and only describes communication pathways, and may include entities not in $\mathcal{G}_d$ (e.g. sensor measurements). Thus, \eqref{eq:distributed_requirement} requires each control input to depend only on local neighborhood information with accessibility defined by $\mathcal{G}_c$. We note that the stability guarantee relies on the existence of a model and decentralized stabilizing controller for the system. However, this stands as a mild assumption in many real-world settings where such models and controllers exist.

We aim to synthesize by-design stabilizing and distributed policies that maximize the discounted cumulative reward in \eqref{rl_formluation}, while generalizing to larger instances of the problem without requiring re-training.

\subsection{The Benefits of GNNs for Distributed Control}
A natural way to enforce the distributed controller structure in \eqref{eq:distributed_requirement} is to unify the policies $\pi_\theta^{[i]}$ using a Graph Neural Network (GNN). A GNN $\Phi_\theta: \mathbb{R}^{|\mathcal{V}| \times F_\text{in}} \rightarrow \mathbb{R}^{|\mathcal{V}| \times F_\text{out}}$ consists of $L$ layers of neighborhood level computations on a given graph $\mathcal{G} = (\mathcal{V}, \mathcal{E})$, mapping node features $x^{[i]} \in \mathbb{R}^{F_\text{in}}, i=1,...,|V|$ to embeddings of size $F_\text{out}$. A layer $l$ performs 2 core operations:
\begin{enumerate}
    \item \textbf{Message Computation:} For each node $v \in \mathcal{V}$ with associated hidden embeddings $h_l^{[i]} \in \mathbb{R}^{F_l}$, $i=1,...,|\mathcal{V|}$, produce \textit{messages} through a function $g_\text{msg} : \mathbb{R}^{F^l} \rightarrow \mathbb{R}^{F^{l,m}}$ mapping $\{h_l^{[i]}\}_{i=1}^{|\mathcal{V}|} \rightarrow \{m_l^{[i]}\}_{i=1}^{|\mathcal{V}|}$, where $F^{l,m}$ denotes the dimension of the messages.
    \item \textbf{Message Aggregation:} For each neighborhood $\mathcal{N}_i$, \textit{aggregate} the computed messages using a function $g_\text{agg}: ({\mathbb{R}^{F^{l,m}}})^{|\mathcal{N}_i|}\rightarrow \mathbb{R}^{F^{l+1}}$, which combines the messages from each neighborhood to produce the subsequent set of hidden embeddings $\{h_{l+1}^{[i]}\}_{i=1}^{|\mathcal{V}|}$ for layer $l+1$.
\end{enumerate}
To make  our setup computationally compatible with this framework, we modify the global system state \eqref{eq:nonlinsys} to make it \textit{horizontally} stacked, denoted as $X_t = \mathrm{col}_{i \in V_c}(x_t^{[i]})^T \in \mathbb{R}^{|\mathcal{V}_c| \times \max_i \{n_i\}}$, where $\max_i\{n_i\}$ is the largest subsystem dimension. This structure compacts the node information in the graph into a single matrix, which enables efficient graph-level computations. To encode the structure of $\mathcal{G}_c = (\mathcal{V}_c, \mathcal{E}_c)$, we denote by $S_c \in \mathbb{R}^{|\mathcal{V}_c|\times |\mathcal{V}_c|}$ a support matrix consistent with the sparsity of $\mathcal{G}_c$ (e.g the adjacency matrix). Given this definition, the message computation and aggregation for a given layer are combined in a single step 
\begin{align}
    H^{l} = \sigma\left ( G(H^{l-1};S_c, \theta^{l-1}) \right ),
\end{align}
where $\sigma$ is a nonlinearity, $H^l = \text{col}_{i\in \mathcal{V}_c}(h_l^{[i]})^T$, $\theta^l$ are the $l$-layer parameters, and $G$ is a function that combines the message computation and passing operations. As an example,
\begin{align}
    G(H^l; S_c, \theta^l) = D^{-1}S_cH^lW^l
\end{align}
for a Graph Convolutional Network \citep{kipf2016semi}, where $D = \text{diag}(|\mathcal{N}_1|,...,|\mathcal{N}_{|\mathcal{V}|}|)$, $S_c$ is the adjacency matrix, and $\theta^l = W^l$ denotes the weight matrix. A GNN $\Phi_\theta$ is a cascade of such layers. Using an L-layer GNN, we embed the global state as 
\begin{align}
    &Y_t = \Phi_\theta(X_t; S_c)
\end{align}
where $Y_t \in \mathbb{R}^{|\mathcal{V}| \times F_\text{out}}$ are the computed state embeddings. 

This formulation has several desirable properties for distributed control setups. First, GNNs are permutation equivariant, satisfying
\begin{align}\label{perm_eq}
    P\Phi_\theta(X;S) = \Phi_\theta(PX; PSP^T),
\end{align}
where $P$ is a permutation matrix. In other words, changing the node order will reorder the output matrix accordingly. Furthermore, GNNs are also Lipschitz with respect to changes in the graph topology \citep{gama2020stability}. 

 These attributes lends themselves well for distributed control. Permutation equivariance allows GNN policies to generalize across different system sizes, while the Lipschitz property makes the closed-loop behavior of GNN policies robust to time-varying communication topologies. However, using GNNs directly as closed-loop policies can induce instability unless the parameters are carefully regularized  \citep{gama2022distributed}. This conservatism presents a trade-off between stability and expressiveness. We now propose a parameterization that guarantees stability irrespective of the chosen model parameters, enabling stable yet GNN expressive policies.

\section{Methodology}

To solve \eqref{rl_formluation}, we introduce an expressive class of stable policies based on \cite{furieri2025mad}, extending its centralized and deterministic computations to the distributed and stochastic setting. We present a decomposition of the controller into a magnitude and direction term using GNNs, and prove its robustness against changes in both the graph communication topology and the model parameters.

\subsection{Separating Magnitude and Direction}
To restrict our search over the space of stabilizing policies, we recall the following sufficient and necessary conditions from \cite{furieri2025mad}:

\begin{theorem}\label{theorem_1}
\citep{furieri2025mad} Let $\boldsymbol{\mathcal{F}}, \boldsymbol{\mathcal{M}} \in \mathcal{L}_p$ and $\boldsymbol{\mathcal{M}}: \ell^n \rightarrow \ell^m$ be a causal operator. Then the closed-loop system induced by
\begin{align}\label{policy_thm1}
    \mathbf{u} = \boldsymbol{\mathcal{M}}(\mathbf{x} - \mathbf{F}(\mathbf{x}, \mathbf{u})) = \boldsymbol{\mathcal{M}}(\mathbf{w})
\end{align}
satisfies $\mathbf{w} \rightarrow (\mathbf{\mathbf{x}, \mathbf{u}}) \in \mathcal{L}_p$. On the other hand, if there exists a causal controller $\mathbf{u} = \boldsymbol{\pi}(\mathbf{x})$ satisfying $\mathbf{w} \rightarrow (\mathbf{\mathbf{x}, \mathbf{u}}) \in \mathcal{L}_p$, then there exists an operator $\boldsymbol{\mathcal{M}} \in \mathcal{L}_p$ which induces the same closed-loop behavior.
\end{theorem}

Theorem \ref{theorem_1} states that the task of searching over stabilizing policies is equivalent to searching over operators in $\mathcal{L}_p$ that map disturbances to control inputs. Thus, \eqref{rl_formluation} can be solved by parameterizing $\mathcal{L}_p$ operators $\boldsymbol{\mathcal{M}}$ to automatically satisfy the stability constraint. However, deriving $\mathbf{u}$ purely from disturbances leaves the entire closed-loop behavior to be characterized by how well one can approximate the space of $\mathcal{L}_p$ operators.

A notable limitation of the policy under Theorem \ref{theorem_1} is that it is \emph{not} necessarily distributed. Works such as \cite{saccani2024optimal} consider a distributed variant of the theorem, but still relies purely on disturbance feedback. \cite{furieri2025mad} introduces an extension of the policy class in \eqref{policy_thm1}, which introduces direct state-feedback while preserving closed-loop stability and completeness of the parametrization of stabilizing policies. This makes the setup compatible with modern RL pipelines. In this work, we turn our attention to distributed stabilizing policies that are compatible with multi-agent RL. Specifically, we unite the properties of the aforementioned approaches with the benefits of GNNs to synthesize a policy that is both stabilizing, scalable and expressive enough to achieve state-of-the-art performance. Our proposed policy embeds GNNs in a stochastic version of MAD policies in \cite{furieri2025mad}. Specifically, our proposed parameterization is as follows:
\begin{align}\label{policy_param}
    &u_t = |\mathcal{M}_t(w_{t:0})|\cdot D_t, \notag \\
    &D_t \sim \mathcal{P}(\cdot| \psi( x_{t:0})), \notag \\
    &|D_t| \leq 1, \boldsymbol{\mathcal{M}} \in \mathcal{L}_p,
\end{align}
where $\boldsymbol{\mathcal{M}}$ must be constructed as a stable GNN operator. The direction term $D_t$ is sampled from a distribution $\mathcal{P}$ conditioned on the output of a function $\psi$ of the past states $x_{t:0}$. We define
\begin{align}
    &\mathcal{M}_t(w_{t:0}) = \text{LRU}(z_{t:0}),
\end{align}
where $\text{LRU}$ is a \textit{Linear Recurrent Unit} acting node wise on each individual $z_t^{[i]}$, defined as
\begin{align}
    &\xi_{t+1}^{[i]} = \Lambda \xi_t^{[i]} + \Gamma(\Lambda)Bz_t^{[i]}, \notag \\
    &\text{LRU}(z_{t:0}^{[i]}) = \text{NN}(\Re(C\xi_t^{[i]}) + Dz_t^{[i]}; \phi) + Fz_t^{[i]},
\end{align}
where NN denotes a neural network, $\Re$ is the real operator and $\Lambda, B, C, D, F, \phi$ are learnable parameters. We enforce stability by designing $\Lambda$ such that its eigenvalues $|\lambda_i| < 1$, with $\Gamma(\Lambda)$ being a diagonal normalization term. Lastly, the inputs to the LRU are computed using a GNN
\begin{align}
    z_t = \Phi(W_t;S_c, \theta_1),
\end{align}
where $W_t$ is the horizontally stacked disturbances observed at time $t$. For the direction term $D_t$, we use a separate GNN to define
\begin{align}\label{direction_output}
    &\psi(x_{t:0}) = \text{RNN}(v_{t:0}, h_{t:0}), \text{ } v_t = \Phi(X_t;S_c, \theta_2),
\end{align} 
where RNN denotes a Recurrent Neural Network \citep{elman1990finding}, a type of neural architecture with a running sequence dependent memory $h_{t:0}$ and external input $v_t$ at each time step. The RNN is applied to each node on the sequence of past GNN outputs $v_{t:0}^{[i]}$ and hidden state $h_{t:0}^{[i]}$. Note that the RNN and LRU make up the internal controller states $s_{t:0}^{[i]} = (\xi_{t:0}^{[i]}, h_{t:0}^{[i]})$.

\begin{corollary}\label{coro_1}
    Under Assumption \ref{ass:open_loop stability}, a distributed controller defined according to \eqref{policy_param}-\eqref{direction_output} induces a mapping $\mathbf{w} \rightarrow (\mathbf{x}, \mathbf{u}) \in \mathcal{L}_p$ if $\Phi(\cdot;S_c,\theta_1)$ has finite $\mathcal{L}_p$-gain $\gamma(\Phi) < \infty$.
\end{corollary}
\textbf{Proof.} By assumption, we have  
\begin{align}
    ||\mathbf{z}|| \leq \gamma(\Phi)||\mathbf{w||}.
\end{align}
Since $\Lambda$ is chosen as a diagonal matrix with eigenvalues $\lambda_i < 1$, we have $\gamma(\text{LRU}) < \infty$. Thus, we can write
\begin{align}
    ||\boldsymbol{\mathcal{M}}(\mathbf{w})|| &= ||\text{LRU}(\mathbf{z})|| \leq \gamma(\text{LRU})||\mathbf{z}|| \notag \\
    &\leq \gamma(\text{LRU})\gamma(\Phi)||\mathbf{w||}.
\end{align}
Next, we have
\begin{align}
    ||\mathbf{u}|| &= ||\boldsymbol{\mathcal{M}}(\mathbf{w})\cdot\mathbf{D}|| \leq ||\boldsymbol{\mathcal{M}}(\mathbf{w})|| \notag \\
    &\leq \gamma(\text{LRU})\gamma(\Phi)||\mathbf{w||}.
\end{align}
The first inequality holds due to the constraint $|D_t| \leq 1$, therefore $|\mathcal{M}_t(w_{t:0})\cdot D_t| \leq |\mathcal{M}_t(w_{t:0})|$ for all $t$. Therefore, $\mathbf{u} \in \mathcal{L}_p$. By Theorem \ref{theorem_1}, $\mathbf{w} \rightarrow (\mathbf{u}, \mathbf{x}) \in \mathcal{L}_p$. 
\hfill $\blacksquare$

 As shown in Corollary \ref{coro_1}, our parameterization guarantees stability of the closed-loop system. This construction decouples the handling of stability and expressivity, allowing the magnitude term to ensure stability while leveraging GNNs to freely control the direction to produce rich closed-loop behaviors. 
% \begin{remark}
%     While the control input computations are distributed, each node might still receive information from its $n$-hop neighbors. For example, a L-layer GNN performs L local information aggregation operations, meaning the final control uses information from its L-hop neighbors.
% \end{remark}
\subsection{Robustness Against Perturbations}
Next, we show the robustness properties of the policy parameterization \eqref{policy_param} against perturbations in the graph communication topology and model weights for a specific GNN architecture. We follow \cite{nayak2023scalable} and use the Unified Message Passing Model (UniMP) presented in \cite{shi2020masked}. The forward pass is defined as
\begin{align}\label{gnn_forward}
    H^{l} = \sigma(\tilde{A}H^{l-1}W^{l-1} + H^{l-1}B^{l-1}).
\end{align}
In \eqref{gnn_forward}, $\sigma$ is nonlinear activation such that $|\sigma(x)| \leq L_\sigma|x|$ and $\sigma(0) = 0$, $H^l \in \mathbb{R}^{N \times F^l}$ is the $l$-th layer hidden state with $N$ denoting the number of nodes, and $F^l$ the embedding dimension at layer $l$. The hidden states are transformed via learned weights and biases $W^{l-1} \in \mathbb{R}^{F^{l-1} \times F^{l}}$ and $B^{l-1} \in \mathbb{R}^{F^{l-1} \times F^l}$ at each layer. The matrix $\tilde{A} \in \mathbb{R}^{N \times N}$ has a sparsity pattern that encodes the attention scores. We refer the reader to \cite{shi2020masked} for a full characterization of this architecture. For simplicity, we consider the update \eqref{gnn_forward} with the attention matrix replaced by a generic support matrix $S$.
\begin{lemma}\label{lemma_1}
    Let $\Phi, \hat{\Phi}$ be two $L$-layer GNNs with updates
    \begin{align}\label{generic_gnn}
            H^{l} = \sigma(SH^{l-1}W^{l-1} + H^{l-1}B^{l-1}),
    \end{align}
    with support matrices and parameters $(S, W, B) $ and $ (\hat{S}, \hat{W}, \hat{B})$. Denote the differences between these values as $(\Delta S, \Delta W, \Delta B)$, i.e $\hat{S} = S + \Delta S, \hat{B} = B + \Delta B, \hat{W} = W + \Delta W$. Let $|| \cdot ||$ denote the element-wise matrix $p$ norm. Then for a given input $X$, the output differences between the GNNs are bounded by
    \begin{align}
        || \Phi(X) - \hat{\Phi}(X) || \leq L_\sigma^L ||X|| \left( \sum_{i=0}^{L-1}\Delta_i \prod_{j=i+1}^{L-1}\rho_j \prod_{v=0}^{i-1}\zeta_v \right) 
    \end{align}
where $\Delta_{i} = ||\Delta S|| \cdot||\Delta W^i|| + ||\Delta S|| \cdot ||W^i|| + ||\Delta B^i|| + ||S|| \cdot ||\Delta W^i||$, $\zeta_v= ||\hat{S}||\cdot||\hat{W}^v|| + ||\hat{B}^v||$ and  $\rho_j= ||S||\cdot||W^j|| + ||B^j||$.
\end{lemma}
\textbf{Proof.} We begin by unpacking the GNN computations
 \begin{align}\label{gnn_proof_1}
        &|| \mathbf{\Phi}(X) - \hat{\mathbf{\Phi}}(X) || = ||H^L - \hat{H}^L|| = \notag \\
        &=|| \sigma(SH^{L-1}W^{L-1} + H^{L-1}B^{L-1}) - \notag \\
        &-\sigma(\hat{S}\hat{H}^{L-1}\hat{W}^{L-1} + \hat{H}^{L-1}\hat{B}^{L-1}) || \leq \notag \\
        &\leq L_\sigma || SH^{L-1}W^{L-1} + H^{L-1}B^{L-1} - \notag \\
        &-\hat{S}\hat{H}^{L-1}\hat{W}^{L-1} + \hat{H}^{L-1}\hat{B}^{L-1} || \leq \notag \\
        &\leq L_\sigma (||S||\cdot||W^{L-1}|| + ||B^{L-1}||) \cdot ||H^{L-1} - \hat{H}^{L-1}|| + \notag \\
        &+L_\sigma(||\Delta S||\cdot ||\Delta W^{L-1}|| + ||\Delta B^{L-1}||) \cdot||\hat{H}^{L-1}|| + \notag \\
        &+ L_\sigma ||\Delta S|| \cdot ||W^{L-1}|| \cdot |\hat{H}^{L-1}|| = \notag \\
        &= L_\sigma \left( ||S|| \cdot||W^{L-1}|| + ||B^{L-1}|| \right) \cdot||H^{L-1} - \hat{H}^{L-1}|| + \notag \\
        &+ L_\sigma ( ||\Delta S|| \cdot ||\Delta W^{L-1}|| + ||\Delta B^{L-1}|| + ||\Delta S|| \cdot ||W^{L-1}|| + \notag \\ 
        &+||S|| \cdot ||\Delta W^{L-1}|| ) \cdot ||\hat{H}^{L-1}||. 
    \end{align}
    
    Let $\Delta_{L-1} = ||\Delta S|| \cdot ||\Delta W^{L-1}|| + ||\Delta B^{L-1}|| + ||\Delta S|| \cdot ||W^{L-1}|| + ||S|| \cdot ||\Delta W^{L-1}||$ and $\rho_{L-1} = ||S|| \cdot||W^{L-1}|| + ||B^{L-1}||$, then \eqref{gnn_proof_1} can be written as the recurrent relation
    \begin{align}\label{gnn_bound_1}
        &|| \mathbf{\Phi}(X) - \hat{\mathbf{\Phi}}(X) || \leq \notag \\
        &\leq L_\sigma\rho_{L-1} || H^{L-1} - \hat{H}^{L-1} || + L_\sigma \Delta_{L-1} ||\hat{H}^{L-1}||.
    \end{align}

    The norm of $\hat{H}^{L-1}$ can be bounded as
    \begin{align}
        &||\hat{H}^{L-1}|| = ||\sigma(\hat{S}\hat{H}^{L-2}\hat{W}^{L-2} + \hat{H}^{L-2}\hat{B}^{L-2})|| \leq \notag \\
        & \leq L_\sigma||\hat{S}\hat{H}^{L-2}\hat{W}^{L-2} + \hat{H}^{L-2}\hat{B}^{L-2}|| \leq \notag \\
        &\leq L_\sigma \left ( ||\hat{S}|| \cdot ||\hat{W}^{L-2}|| + ||\hat{B}^{L-2}|| \right ) \cdot ||\hat{H}^{L-2}|| = \notag \\
        &=L_\sigma \zeta_{L-2} ||\hat{H}^{L-2}||,
    \end{align}
    where $\zeta_{L-2} = ||\hat{S}|| \cdot ||\hat{W}^{L-2}|| + ||\hat{B}^{L-2}||$. Writing this out recursively, we get
    \begin{align}\label{H_hat_bound}
        ||\hat{H}^{L-1}|| \leq L_\sigma^{L-1}||X|| \prod_{k=0}^{L-2}\zeta_k.
    \end{align}

Plugging \eqref{H_hat_bound} into \eqref{gnn_bound_1}, we get
\begin{align}
    &|| \mathbf{\Phi}(X) - \hat{\mathbf{\Phi}}(X) || \leq \notag \\
    &\leq L_\sigma\rho_{L-1} || H^{L-1} - \hat{H}^{L-1} || + L_\sigma^L \Delta_{L-1}||X|| \prod_{k=0}^{L-2}\zeta_k.
\end{align}
Unrolling the recursive relation from $l=1,...,L$, we get the following pattern:
\begin{align}
    &l=1\text{: } ||H^1 - \hat{H}^1|| \leq L_\sigma \Delta_0 ||X||, \notag \\
    &l=2\text{: } ||H^2 - \hat{H}^2|| \leq L_\sigma^2 ||X||\left( \Delta_0 \rho_1 + \zeta_0 \Delta_1 \right), \notag \\
    &l=3\text{: } ||H^3 - \hat{H}^3|| \leq L_\sigma^3 ||X|| \left( \Delta_0 \rho_1 \rho_2 + \zeta_0 \Delta_1 \rho_2 + \Delta_2 \zeta_0 \zeta_1 \right), \notag \\
    &\vdots \notag \\
    &l=L\text{: } ||H^L - \hat{H}^L|| \leq L_\sigma^L ||X|| \left( \sum_{i=0}^{L-1}\Delta_i \prod_{j=i+1}^{L-1}\rho_j \prod_{v=0}^{i-1}\zeta_v \right). \notag
\end{align}
\hfill $\blacksquare$

Lemma \ref{lemma_1} leads us to the following result about closed-loop robustness.
\begin{theorem}\label{theorem_2}
    Let $\mathbf{\Phi}, \hat{\mathbf{\Phi}}$ be two L-layer GNN operators with updates specified by \eqref{generic_gnn}, with support matrices and parameters differing by $(\Delta S, \Delta W, \Delta B)$. Assume $\boldsymbol{\mathcal{F}}$ is a causal operator with finite $\ell_p$-gain $\gamma(\boldsymbol{\mathcal{F}})$. Then for all disturbances $\mathbf{w} \in \ell_p$, the deviations in closed-loop trajectories induced by $\mathbf{\Phi}$ and $\hat{\mathbf{\Phi}}$ when used in the policy \eqref{policy_param}-\eqref{direction_output} are upper bounded by
    \begin{align}\label{traj_bound}
        &||\mathbf{x} - \hat{\mathbf{x}}|| \leq \notag \\
       &\leq \gamma(\boldsymbol{\mathcal{F}}) \gamma(\text{LRU})L_\sigma^L ||\mathbf{w}|| \left( \sum_{i=0}^{L-1}\Delta_i \prod_{j=i+1}^{L-1}\rho_j \prod_{v=0}^{i-1}\zeta_v + 2\prod_{k=0}^{L-1}\zeta_k \right)
    \end{align}
\end{theorem}
\textbf{Proof.} We begin leveraging the assumption $\boldsymbol{\mathcal{F}} \in \mathcal{L}_p$ with gain $\gamma(\boldsymbol{\mathcal{F}})$. We can write
\begin{align}\label{x_diff_bound}
    &||\mathbf{x} - \hat{\mathbf{x}}|| = ||\boldsymbol{\mathcal{F}}(\mathbf{u}, \mathbf{w}) - \boldsymbol{\mathcal{F}}(\hat{\mathbf{u}}, \mathbf{w})|| \leq \notag \\
    &\leq \gamma(\boldsymbol{\mathcal{F}})||(\mathbf{u}, \mathbf{w}) - (\hat{\mathbf{u}}, \mathbf{w})|| = \gamma(\boldsymbol{\mathcal{F}})||\mathbf{u} - \hat{\mathbf{u}}||.
\end{align}
For each time $t$, the differences between control inputs are
\begin{align}
    &u_t - \hat{u}_t = |\mathcal{M}_t(w_{t:0})|\cdot D_t - |\hat{\mathcal{M}_t}(w_{t:0})|\cdot \hat{D}_t = \notag \\
    &= (|\mathcal{M}_t(w_{t:0})| - |\hat{\mathcal{M}}_t(w_{t:0})|)\cdot D_t + |\hat{\mathcal{M}}_t(w_{t:0})|\cdot(D_t - \hat{D}_t).
\end{align}
Therefore, it holds that
\begin{align}
    |u_t - \hat{u}_t| &\leq |\mathcal{M}_t(w_{t:0})| - |\hat{\mathcal{M}}_t(w_{t:0})| + 2|\hat{\mathcal{M}}_t(w_{t:0})| \leq \notag \\
    &\leq |\mathcal{M}_t(w_{t:0}) - \hat{\mathcal{M}}_t(w_{t:0})| + 2|\hat{\mathcal{M}}_t(w_{t:0})|,
\end{align}
where we utilized that $|D_t| \leq 1$. This shows that we can write
\begin{align}
    ||\mathbf{u} - \hat{\mathbf{u}}|| \leq ||\boldsymbol{\mathcal{M}}(\mathbf{w}) - \hat{\boldsymbol{\mathcal{M}}}(\mathbf{w})|| + 2||\hat{\boldsymbol{\mathcal{M}}}(\mathbf{w})||.
\end{align}
Using our policy parameterization \eqref{policy_param}-\eqref{direction_output}, we have $\boldsymbol{\mathcal{M}}(\mathbf{w}) = \text{LRU}(\mathbf{\Phi}(\mathbf{w}))$, giving us
\begin{align}\label{u_diff_bound}
    ||\mathbf{u} - \hat{\mathbf{u}}|| \leq \gamma(\text{LRU})|| \mathbf{\Phi}(\mathbf{w}) - \hat{\mathbf{\Phi}}(\mathbf{w}) || + 2\gamma(\text{LRU})|| \hat{\mathbf{\Phi}}(\mathbf{w})||.
\end{align}
Using Lemma \ref{lemma_1}, we have
\begin{align}\label{phi_diff_bound}
    || \mathbf{\Phi}(\mathbf{w}) - \hat{\mathbf{\Phi}}(\mathbf{w}) || \leq L_\sigma^L ||\mathbf{w}|| \left( \sum_{i=0}^{L-1}\Delta_i \prod_{j=i+1}^{L-1}\rho_j \prod_{v=0}^{i-1}\zeta_v \right).
\end{align}

Using the definition of the $\ell_p$-gain of an $L$-layer GNN from \eqref{H_hat_bound} and inserting \eqref{phi_diff_bound} into \eqref{u_diff_bound}, then porting the result to \eqref{x_diff_bound} and factorizing, we arrive at
\begin{align}
        &||\mathbf{x} - \hat{\mathbf{x}}|| \leq \notag \\
       &\leq \gamma(\boldsymbol{\mathcal{F}}) \gamma(\text{LRU})L_\sigma^L ||\mathbf{w}|| \left( \sum_{i=0}^{L-1}\Delta_i \prod_{j=i+1}^{L-1}\rho_j \prod_{v=0}^{i-1}\zeta_v + 2\prod_{k=0}^{L-1}\zeta_k \right).\notag
\end{align}
\hfill $\blacksquare$

Theorem \ref{theorem_2} implies that the policy parameterization \eqref{policy_param}-\eqref{direction_output} is robust to changing communication topologies and model parameters (since $\tilde{A}$ is a support matrix), and that the trajectory deviations scales with the size of the perturbations. Finite changes in the graph structure and weights can therefore never cause the closed-loop system to be unstable, even if the policy is untrained. This allows stability to be always retained, while enabling generalization of performance to unseen topologies without the risk of destabilization. We note that the bound in Theorem \ref{theorem_2} is not tight, as it would be non-zero even in the absence of perturbations. This is caused by the stochasticity of $D_t$, which induces trajectory mismatches even under identical model parameters. However, the benefits of stochastic policies outweigh the lack of tightness, and the trajectories still eventually coincide if $\mathbf{w} \in \ell_p$.

\subsection{Practical Implementation with PPO}
When training stochastic RL policies, it is often desirable to prevent updates that result in overly large changes in the policy to improve training stability. Proximal Policy Optimization (PPO) \citep{schulman2017proximal} is a popular policy-gradient method that encourages this behavior by constraining the update ratio $\frac{\pi_\theta(u_t | x_t)}{\pi_{\theta_\text{old}}(u_t | x_t)}$ between the new and old policies. To make our policy compatible, we need to compute the log probabilities of the update ratio between the new and old policies. To simplify notation, we abstract the total control as the simplified representation
\begin{align}
    &u_t = u_{\text{base},t} + |\mathcal{M}_t(w_{t:0})| \cdot g(D_t)
\end{align}
with $D_t \sim \mathcal{P}(d | x_{t:0})$, $u_{\text{base},t}$ is the stabilizing base controller, and $g: \mathbb{R}^m \rightarrow \mathbb{R}^m$ is an invertible function ensuring $|D_t| \leq 1$ (e.g tanh). Let $z_t = \frac{u_t - u_{\text{base},t}}{|\mathcal{M}_t(w_{t:0})|}$, then the policy describing the distribution of $u_t$ is
\begin{align}
    &\pi(u_t | x_{t:0}) = \mathcal{P}(D_t | x_{t:0}) \left| \frac{d D}{du} \right| = \mathcal{P} (g^{-1}(z_t)|x_{t:0}) \left| \frac{dg^{-1}}{dz_t} \frac{dz_t}{du_t}\right | \notag \\
    &=\mathcal{P}( g^{-1}(z_t) | x_{t:0}) \frac{1}{|\mathcal{M}_t(w_{t:0})|} \left | \frac{dg^{-1}}{dz_t} \right |.
\end{align}
This leads us to the following computation:
\begin{align}
    &\log \pi(u_t | x_{t:0}) = \log \mathcal{P}(g^{-1}(z_t) | x_{t:0}) - \log |\mathcal{M}_t(w_{t:0})| + \notag \\
    &+\log \left | \frac{dg^{-1}}{dz_t} \right |.
\end{align}

In other words, the computation of the distribution over control inputs requires reconstructing the direction term $D_t$ which was sampled during the roll-out. Another takeaway is that one would also need to collect $w_t$ and base controller inputs $u_{\text{base},t}$ during the roll-outs apart from the states and control inputs. 
 
\section{Numerical Experiments}
In this section, we describe the details of our experiments. Our goal is to highlight how the deliberately chosen set of ingredients that make up our approach enable synthesis of scalable and generalizable controllers, while being expressive and closed-loop stable by design. We demonstrate this through careful ablations of our method, individually removing select components to isolate the impact of their absence. Specifically, we study the following ablations: 1) Setting $\mathcal{M}_t = 1$ and removing the stabilizing base policy to study the absence of the stability guarantees (equivalent to \cite{nayak2023scalable}), 2) Parameterize $\psi$ with a multi-layer perceptron (MLP) taking \textit{centralized} state inputs to examine the value of the GNN based policy, 3) Replace the GNN-based critic network during training with an MLP to test the influence of a graph-based critic while keeping the actor consistent with \eqref{policy_param}. Our code is available in our repository \footnote{https://github.com/Mudhdhoo/Scaling-Up-Stability}.

\begin{figure}[!t]
    \centering
    \includegraphics[width=\linewidth]{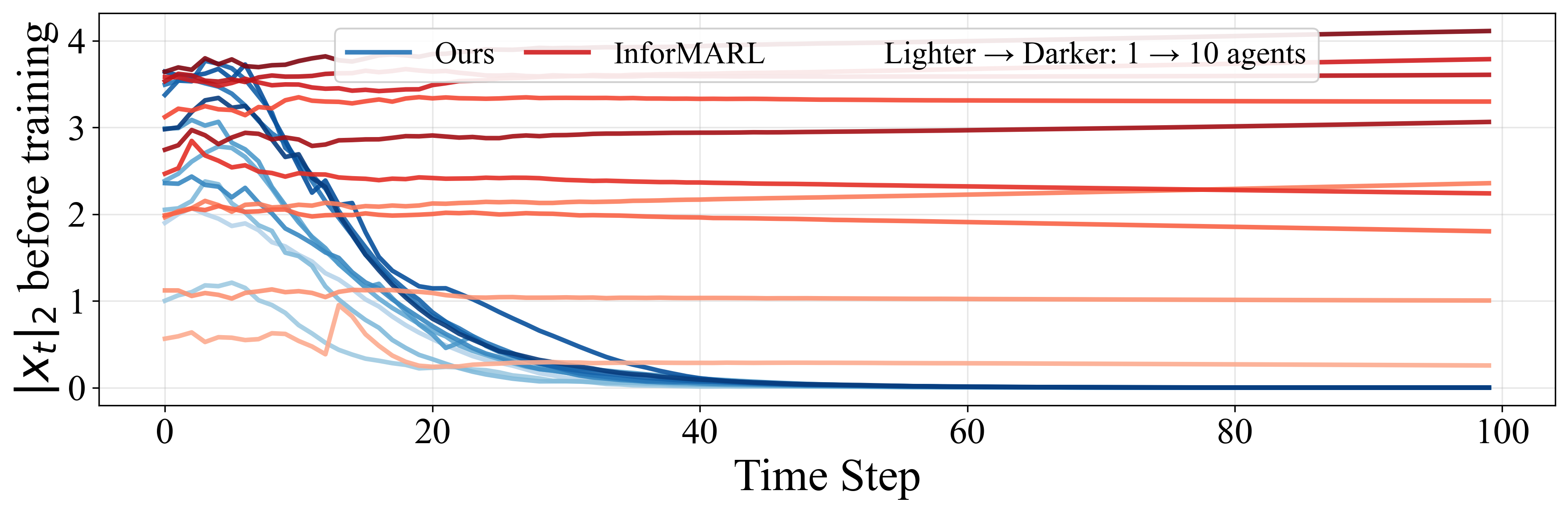}
    \includegraphics[width=\linewidth]{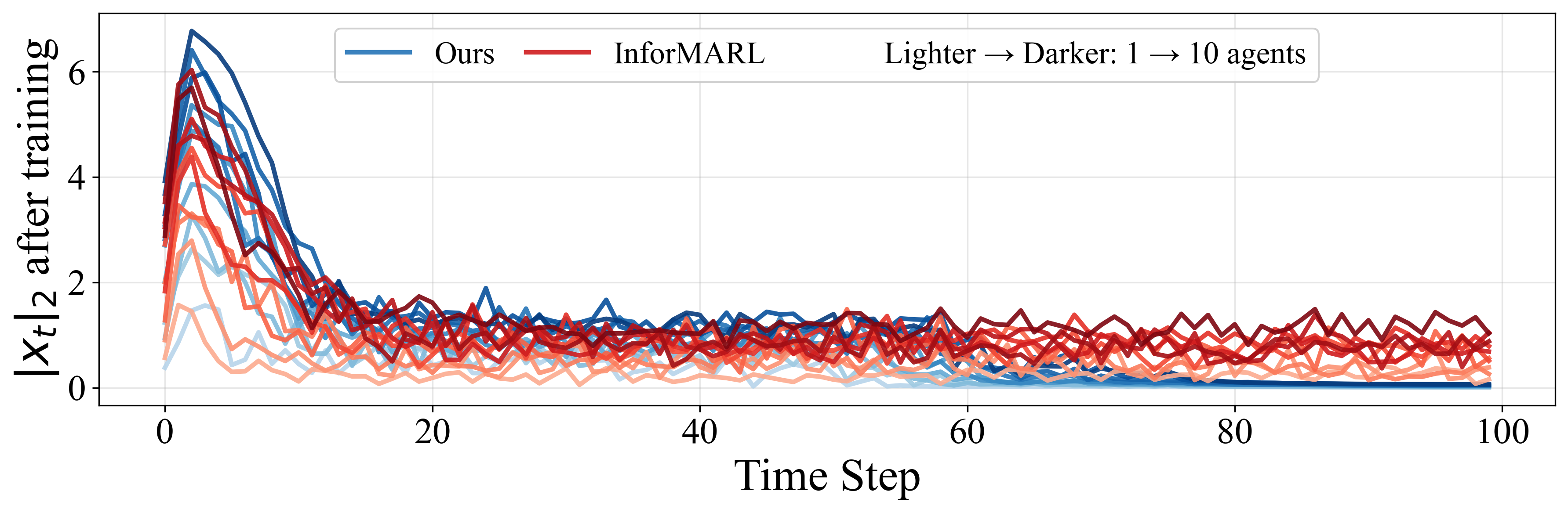}
    \caption{Evolution of the state norms across 10 different runs, with the number of agents ranging from 1 to 10. Lighter color denotes fewer agents, while darker denotes more. Top shows untrained policies, bottom shows trained ones.}
    \label{fig:stability_comp}
\end{figure}

\subsection{Environment and Setup}
We benchmark on the Multi-Agent Particle Environment used by \cite{nayak2023scalable}. It consists of $N$ agents modeled as 2D integrators with nonlinear collision and speed dynamics. The state of agent $i$ is $x_t^{[i]} = [v_{x,t}^{[i]}, v_{y,t}^{[i]}, p_{x,t}^{[i]}, p_{y,t}^{[i]}]$, with $v$ and $p$ denoting the velocity and position of the agent at the $x,y$ coordinates and time $t$. At the start of each episode, every agent is randomly assigned an initial position and a target position $p_{\text{goal}}^{[i]}$. The task is to reach the goal while avoiding collisions with other agents and obstacles. 
\begin{figure}[!ht]
    \centering
    \includegraphics[width=0.8\linewidth]{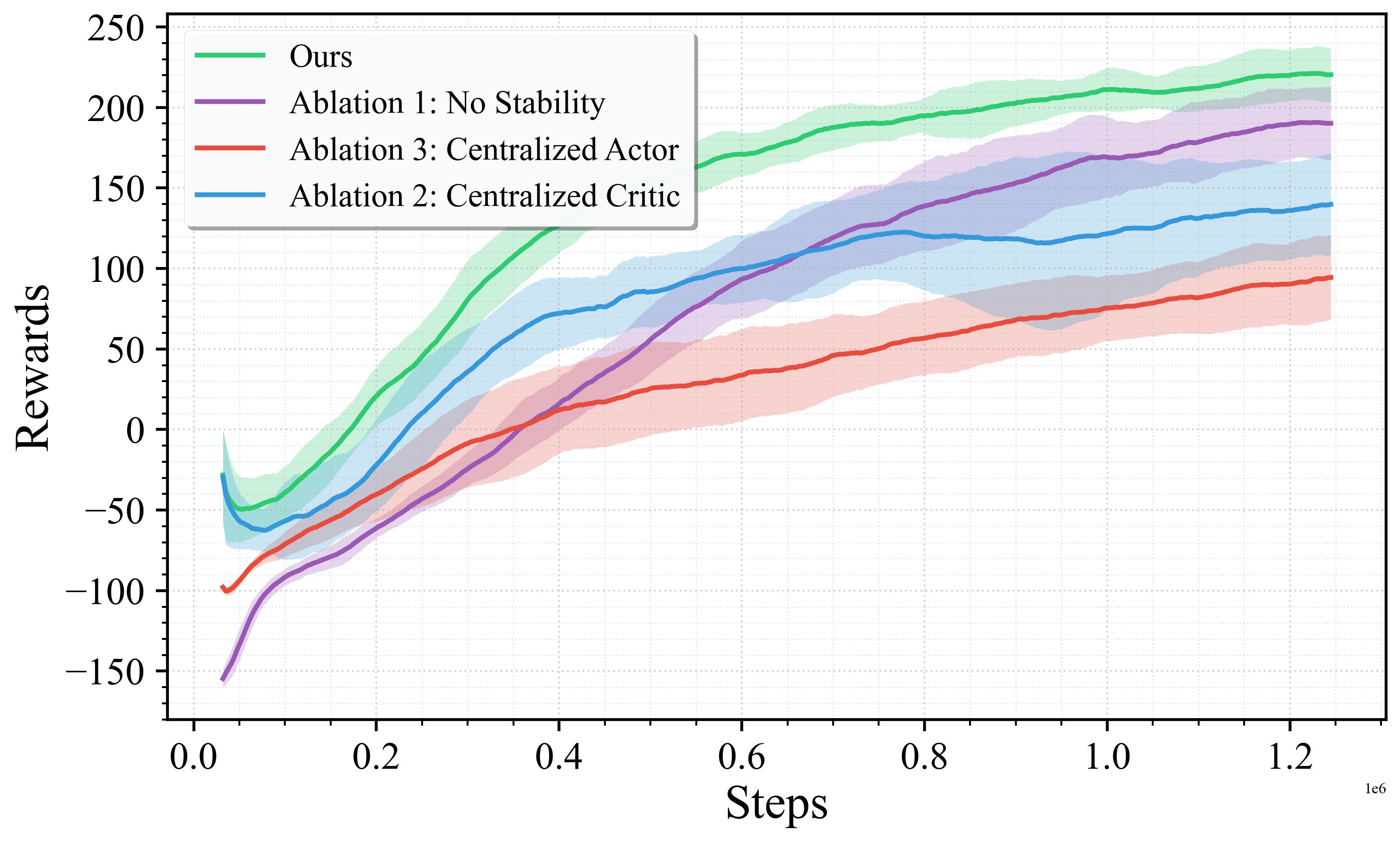}
    \caption{Mean and standard deviation of our method compared to baselines. Trained using PPO on 5 agents across 5 random seeds.}
    \label{fig:training_comp}
\end{figure}
\begin{table}[!ht]
    \centering
    \begin{tabular}{l ccc}
        \toprule
        \diagbox{Ablation}{Num Agents} & N=5 & N=7 & N=10 \\
        \midrule
        Ours &  $\mathbf{274.086}$ & $\mathbf{371.189}$ & $\mathbf{502.914}$  \\
        Ablation 1: No Stability  & 203.512 & 267.172 & 351.562 \\
        Ablation 2: Centralized Actor & 168.535 & N/A & N/A \\ 
        Ablation 3: Centralized Critic & 149.961 & 191.009 & 246.010 \\
        \bottomrule
    \end{tabular}
    \caption{Ablation study: Averaged rewards across 100 seeds and varying number of agents.}
    \label{ablation}
\end{table}
Agents share a global reward $R(x_t, u_t) = \sum_{i=1}^N r(x_t^{[i]}, u_t^{[i]})$, where $r(x_t^{[i]}, u_t^{[i]}) = -\|p_t^{[i]} - p_{\text{goal}}^{[i]}\|_2 + r_{\text{coll},t}^{[i]} + r_{\text{goal},t}^{[i]}$, with 
$p_t^{[i]} = [p_{x,t}^{[i]}, p_{y,t}^{[i]}]$, $r_{\text{coll},t}^{[i]} = -5$ when colliding with an agent or obstacle, and $r_{\text{goal},t}^{[i]} = 5$ upon entering the vicinity of the goal position. Each agent $i$ senses neighbors within a communication radius $r$, receiving disturbances $w_t^{[j]}$ and node features $[x_t^{[j]}, p_\text{goal}^{[j]}, \text{entity}\_\text{type}(j)]$ for all agents $j$ within this radius, where $\text{entity}\_\text{type}(j)$ is a unique index for each entity type. Thus, the GNN operates on each agent’s local, \textit{dynamically} changing communication graph $\mathcal{G}_{c,t}^{[i]}$, with the global graph $\mathcal{G}_{c,t} = \cup_{i}\mathcal{G}_{c,t}^{[i]}$. We implement \eqref{policy_param}-\eqref{direction_output} by choosing $\mathcal{P}$ to be a Gaussian distribution, and let the GNN$\rightarrow$RNN cascade output $\psi(x_{t:0}) = (\mu(x_{t:0}), \Sigma(x_{t:0}))$, the mean and covariance of the distribution. The sample $D_t$ is wrapped around a tanh to ensure $|D_t| < 1$. Pre-stabilization of the system is achieved using a base proportional controller $u_{\text{base},t}^{[i]} = K^{[i]}(p^{[i]}_t - p_{\text{goal}}^{[i]})$ for each agent. We set an upper limit on the value the magnitude can attain, which we found to result in more stable training. All models are trained with around 50k parameters.

% \begin{table}[h]
%     \centering
%     \begin{tabular}{l ccc}
%         \toprule
%         \diagbox{Ablation}{Num Agents} & N=5 & N=7 & N=10 \\
%         \midrule
%         Ours &  $\mathbf{271.215}$ & $\mathbf{380.970}$ & $\mathbf{492.428}$  \\
%         GNN Policy  & 192.812 & 254.364 & 374.741 \\
%         Centralized MAD & 162.003 & N/A & N/A \\
%         \bottomrule
%     \end{tabular}
%     \caption{Ablation study: Averaged rewards across 5 random seeds and varying number of agents.}
%     \label{ablation}
% \end{table}

% \begin{figure}[!t]
%     \centering
%     \includegraphics[width=0.8\linewidth]{figures/reward_comp.png}
%     \caption{Average episode rewards and standard deviations across a varying number of agents for the trained policies from 10 episode rollouts.}
%     \label{fig:rew_comp}
% \end{figure}
\subsection{Results}
 We begin by highlighting the stability properties of our method by comparing it to a purely GNN based policy according to ablation 1, equivalent to InforMARL \citep{nayak2023scalable}. The top of figure \eqref{fig:stability_comp} shows the $\ell_2$ norms of the global states over time when controlled with \emph{untrained} models. The same policy is deployed on a varying number of agents. Despite the absence of training, our method drives the system to equilibrium, while the system remains unstable without our guarantees. The bottom figure shows that after training, the purely GNN based policy drives $x_t$ to oscillate around a neighborhood of the equilibrium without converging, whereas our policy stabilizes the system.

Figure \ref{fig:training_comp} presents the training curves of each individual ablation, showing the mean and standard deviation of accumulated rewards across 5 random training runs in a 5-agent setup. We note that each ablation degrades performance both in terms of rewards achieved and sample efficiency. Notably, the absence of stability guarantees leads to slower convergence and lower average rewards, consistent with previous findings such as in \cite{wang2022learning}. The performance degradation resulting from the use of a centralized actor in place of a GNN highlights the inductive bias of the GNN architecture, which our method effectively leverages. Ablating the GNN-based critic likewise complicates training. However, its training curve still presents itself as competitive to the pure GNN-based policy with a graph-based critic, further suggesting that the stabilizing bias inherited from our parameterization facilitates the learning process.
% Figure \ref{fig:training_comp} presents the training curves of each individual ablation. It shows the mean and standard deviation of accumulated rewards across 5 random training runs in a 5-agent setup. Searching over stabilizing policies allows us to achieve higher average rewards, lower variance, and better sample efficiency, which is consistent with the findings in previous literature such as \cite{wang2022learning}. Interestingly, our method outperforms MAD despite its centralized design, showing that using GNNs to aggregate local information yields richer state representations than simply concatenating the states. This indicates that our parameterization retains expressivity despite the stability constraint, which typically degrades performance. The expressivity stems from the free parameterization of our policy, which allows arbitrary model weights without restricting the search space.

Lastly, table \ref{ablation} compares the test time performance of the ablations through random roll-outs across varying number of agents. Note that unlike GNNs, the centralized MLP actor setup forces a rigid input size due to the stacking of subsystem states instead of aggregating them into single embeddings, making this ablation variant unable to scale across system sizes. The table shows that our proposed combination of GNNs within a MAD like framework leads to improved performance over any ablation on their own. Our approach restricts policy search to a more meaningful space, inheriting the inductive biases and distributed nature of GNNs and the stability and expressivity of the MAD framework.
% Lastly, figure \ref{fig:rew_comp} shows the transferability and robustness of our policy to unseen communication topologies and varying numbers of agents beyond those used during training. The varying number of agents induces communication graphs never seen prior to deployment. The built-in robustness of our parameterization ensures finite perturbation-induced trajectory deviations, and combined with closed-loop stability yields better generalization performance than InforMARL.
\vspace{-0.3cm}
\section{Conclusion}
We proposed a scalable policy class that is expressive, stable by design and enables generalization across system sizes, achieved through combining GNNs with a Youla-like magnitude and direction polar decomposition. Our parameterization yields stabilizing yet expressive controllers that are robust to perturbations in model parameters and communication topology. Ablations on a multi-agent navigation task show built-in stability, scalability and generalization to varying number of agents. Future work includes probabilistic analysis of trajectory deviation bounds and extension of the method to output-feedback systems.
\vspace{-0.2cm}
\bibliography{ifacconf}

\end{document}